\documentclass[reprint,amsmath,amssymb,aps,prc,showpacs]{revtex4-1}
\usepackage{graphicx}   % Include figure files
\usepackage{dcolumn}   % Align table columns on decimal point
\usepackage{bm}        % bold math

\begin{document}

\title{Coulomb effect in Au+Au and Pb+Pb collisions as a function of collision energy}

% The names of the author(s):
\author{D. Cebra, S.G. Brovko, C.E. Flores}
\affiliation{University of California - Davis, Davis, California 95616 }
\author{B.A. Haag}
\affiliation{American River College, Sacramento, California 95841}
\author{J.L. Klay}
\affiliation{California Polytechnic State University, San Luis Obispo, California 93407}

\date{\today}

\begin{abstract}

The subtle differences between positive and negative pion spectra can used
be used to study the nature of the nuclear interaction region in heavy-ion 
collisions. Several large acceptance heavy ion experiments at 
facilities ranging from SIS, the AGS, the SPS, to RHIC have measured 
mid-rapidity $\pi^+$ and $\pi^-$ spectra for central Au+Au or Pb+Pb collisions. 
From these spectra one can create pion ratios as a function of $m_t-m_0$, which 
are used to determine the Coulomb potential, $V_C$, and the 
initial pion ratio, $R_i$, across a range of collision energies from 1 to 
158 AGeV. The implications of the $V_C$ and $R_i$ trends with collision
energy will be discussed.

\end{abstract}

% 25.75.-q ---Relativistic heavy-ion collisions
% 25.75.Dw ---Particle and Resonance production

\maketitle

\section{Introduction}
In heavy-ion collisions, the subtle structure of the pion spectra can be
used to image the positive charge density of the expanding fireball. The
source has a net positive charge from the incident protons in the
projectile and target. The Coulomb force resulting from this net charge
accelerates each charged particle emitted from the source, and in doing so 
changes its final (observed) energy.
Positively charged particles get a modest increase in kinetic energy while
negatively charged particles are reduced. For the purposes of this
paper, it is assumed that the `initial-state' consists of particles that 
are emitted from a thermally equilibrated volume that has reached chemical 
and kinetic freeze-out; the Coulomb interaction is considered to be a 
`final-state'interaction. 
These final-state Coulomb interactions distort the initial
thermal spectra. This distortion can best be observed by comparing spectra
of particles with the same mass but opposite charge. The effect is
greatest for the lightest particles, the pions. The magnitude of the boost
or reduction in kinetic energy is proportional to the Coulomb potential, $V_C$, 
which is determined by the charge distribution of the source and the emission 
point of the pion. Thus by studying the details of the pion spectra, one 
can effectively image the charge density of the source at kinetic freeze-out.

A relative enhancement, at low transverse momentum, of the negative pion
yield with respect to that of the positive pions was first observed
in heavy-ion experiments at the 
Bevalac~\cite{Bene79,Wolf79,Naga81,Wolf82}.
These early results were explained as evidence of the positive Coulomb
potential of the source. A similar enhancement of $\pi^-$'s at low $m_t$
($m_t = \sqrt{p_t^2 + m^2}$) was observed in silicon-induced heavy-ion
reactions at the Alternating Gradient Synchrotron 
(AGS)~\cite{Goni94,Vide95}. Early theoretical
work~\cite{Libb79,Bert80,Gyul81,Stoc86,Li95,Osad96} described these 
results in terms of emission from a static Coulomb source. Interpretation of
these results is confounded by the difficulties in determining the impact
parameter, source velocity, and source size in asymmetric heavy-ion
collisions.

Pion spectra and ratios have been measured in symmetric collisions of the 
heaviest nuclei (Au+Au and Pb+Pb) at facilities ranging from 
SIS~\cite{Pelt97,Wagn98,Reis07}, the
AGS~\cite{Ahle96,E802_98_1,E866_00_1,Barr00,Klay03}, 
the SPS~\cite{Bogg96,Reti01,Ross02,WA98_03,WA98_06,NA44_02,NA49_02,NA49_08}, 
to RHIC~\cite{STAR10}. Several groups have analyzed the pion ratios to 
study the effect of the Coulomb potential.
The KaoS group (at SIS) used a static model to analyze their mid-rapidity pion 
ratio data from 1 AGeV
Au+Au collisions. They concluded that, in general, the freeze-out radius is a
function of pion energy with the high energy pions freezing out
first~\cite{Wagn98}. In a second study of the same data, they found 
the `initial' pion ratio, $R_i$, to be $0.515 \pm 0.05$, and the $V_C$ to
be greater than 20 MeV~\cite{Munt98}. A similar analysis of the E866
mid-rapidity 10.8 AGeV Au+Au data~\cite{Ahle96} found  $R_i = 0.83$ and
$V_C = 9 \pm 3$ MeV. This beam energy dependence is not unexpected. 

The overall pion ratio is heavily influenced by the isospin asymmetry at low
energy. As one increases the energy available for pion production, this
ratio should approach unity. The Coulomb potential is a function of both
the charge of the system and the source radius. The net charge of the 
interaction region is determined by the number of participant protons in the 
overlap volume of the projectile and the target and the degree of baryon stopping 
at a given bombarding energy. The observed decrease in $V_C$ with beam energy is 
therefore indicative of either expansion or of a reduction in baryon stopping. 
These trends should be compared to the results from E877, an experiment 
in which pion ratios are studied at beam rapidity and higher. They find
values of the `overall' pion ratio, $R'=1.09 \pm 0.20$ and $V_C = 31 \pm
22$ MeV~\cite{Barr00}. The distinction between the `initial' pion
ratio, $R$, and the `overall' pion ratio $R'$ will be discussed later in
this paper. Mid-rapidity pion ratio data are also available from the top energy 
at the SPS. NA44 has reported pion ratios which showed
evidence of a Coulomb effect~\cite{Bogg96}. WA98 reported pion ratios which 
are consistent with a finite Coulomb potential~\cite{Reti01,Ross02}. They followed up
this study with a more extensive analysis in which they related
the source potential to the freeze-out time~\cite{WA98_03}. NA49
has studied pion ratios in peripheral collisions~\cite{NA49_05}; these
results are explained in terms of isospin effects, time of initial pion
emission, size of the pion source, and the Coulomb force~\cite{Rybiki2007}.

Theoretical analyses were developed to include 
the Jacobian factor, $d^3p_i/d^3p$,~\cite{Baym96}, and the effects of
radial expansion and pion emission time~\cite{Most95,Teis97,Barz97,Barz98}. 
Ref. \cite{Barz97}  concluded that: the SIS data are consistent
with $V_C = 27$ MeV corresponding to a freeze-out radius of 8 fm, the
AGS data are consistent with a radius of 10 fm, and the early
SPS data (NA44) are consistent with a 9 fm radius. 
A full transport model calculation~\cite{Ayal97,Ayal99} was also
applied to these data. In that analysis, the freeze-out radius was also
found to be 10 fm for the AGS and SPS data sets. Pion emission in heavy-ion 
collisions in the region 1 A GeV is investigated in an isospin dependent 
quantum molecular dynamics model in ref~\cite{Feng2009}. This analysis
demonstrates that in this energy range pions are produced mostly through
the $\Delta$ and $^*N$ channels.

In summary, for experimental mid-rapidity data, the Coulomb potential is
observed to decrease with bombarding energy while the pion ratio rises.
The theoretical analyses suggest an increase in
freeze-out radius, which would correspond to a reduction in $V_C$, for
beam energies from 1 to 10.8 AGeV. 

In this paper, we review and analyze a range of available experimental 
data. 
We focus specifically on the low $m_t-m_0$ ($<$ 0.3 GeV/c$^2$) region,
as this is where the Coulomb potential has the biggest influence on the
spectra and the ratios. We specifically have looked for data sets which 
extend to the lowest $m_t$; this allows a more detailed study of the
effect of expansion on the slow pions.
We have not considered data for collision energies below 1 AGeV.
Although the threshold for inelastic nucleon-nucleon collisions is 0.3 GeV, 
the cross sections for the various $N+N \rightarrow N+\pi$ reactions rise
rapidly from 0.3 before saturating at 1.0 GeV~\cite{VerWest82}. Also, we 
will not present results from bombarding energies above 158 AGeV. Although 
there are a wealth of spectra data from RHIC and the LHC, the degree of baryon
stopping is very low at these higher energies, resulting in a very small
Coulomb potential which would only be seen at the lowest end of the $m_t-m_0$
spectra. None of the RHIC or LHC experiments have especially low $m_t-m_0$ 
thresholds, therefore it is not possible to extract meaningful Coulomb 
potentials from those data. 

\section{Results}
Figure 1 shows the transverse mass spectra at mid-rapidity 
for both positive and negative pions from central Au+Au or Pb+Pb 
collisions at beam energies from 1 to 42 AGeV (references are given in
the figure caption). We note that 42 AGeV is the fixed-target equivalent
energy to the $\sqrt{s_{NN}}$ = 9.2 GeV test run at RHIC from which 
$\pi^\pm$ spectra were published
by the STAR Collaboration~\cite{STAR10}. We also note that several energies
which were studied at the SPS are not represented in this figure. 
Although NA49 has published $\pi^-$ spectra at 40, 80, and 158 AGeV~\cite{NA49_02}, 
they have not published $\pi^+$ spectra at these energies. WA98 has published
pion ratios at 158 AGeV, however they too have only shown their $\pi^-$ spectra.
NA44 has shown both $\pi^-$ and $\pi^+$ spectra at 158 AGeV, however their
acceptance slice only allows low $m_t-m_0$ pions away from mid-rapidity.
Although the positive and negative pion spectra are very similar in slope
for $m_t-m_0 > 0.1$ GeV/c$^2$, a clear difference in the shape of the 
respective spectra is evident below 0.1 GeV/c$^2$ for all beam energies.  In
reference~\cite{Klay03}, the pion spectra were fit with the 
superposition of two independent Boltzmann distributions. 
The two contributions were interpreted to approximately represent the pion 
yields coming from $\Delta$ resonance decays (low temperature component) and 
from direct thermal emission of pions (high temperature component). Studies with
RQMD~\cite{Sorg89} supported this general interpretation. The
high temperature parameter was required to be the same for both the negative and
positive pions for each beam energy. However, the low temperature
parameter was determined independently. 

In this paper, we address the details of the low $m_t-m_0$ shapes and 
amplitudes of the $\pi^+$ and $\pi^-$ spectra in terms of the Coulomb 
potential from an expanding source. This interpretation of the pion spectra 
does not contradict the previous studies~\cite{Klay03,Sorg89}. The fact that 
the low $m_t$ pions are predominantly daughters of $\Delta$ resonance 
decays~\cite{Sorg94,Barr95,Pelt97} explains the difference in the
relative yields of the two charges of pions. However, $\Delta$ resonance 
production does not explain the difference in spectral shapes. The Coulomb
interaction modifies the initial spectral shapes as we show in this paper.

\begin{figure}[th]
\includegraphics[scale=0.4]{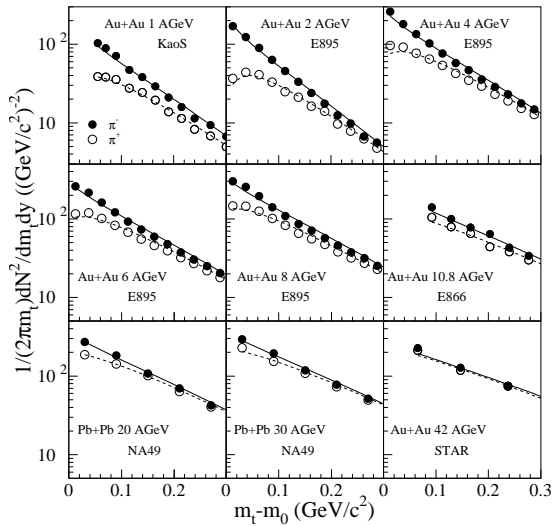}
%\caption{This is a simple caption for $m_t-m_0$ spectra.}
\caption{
The mid-rapidity $m_t-m_0$ spectra are shown for
negative (solid symbols) and positive (open symbols) pions from central Au+Au 
and Pb+Pb collisions
at 1.0~\cite{Wagn98}, 2.0~\cite{Klay03},
4.0~\cite{Klay03}, 6.0~\cite{Klay03}, 8.0~\cite{Klay03}, 10.8~\cite{Ahle96}, 
20~\cite{NA49_08}, 30~\cite{NA49_08}, and 42~\cite{STAR10} AGeV.
The Coulomb distortion is exhibited by the divergence of
the spectra at low $m_t-m_0$. The spectra are fit with a Bose-Einstein
function modified by Coulomb acceleration due to the effective
potential from Eqn.~(4). 
A single temperature parameter is used at each
bombarding energy to simultaneously fit the $\pi^-$ (solid) and the
$\pi^+$ (dashed). }
\label{fig:spec}
\end{figure}

In the analysis outlined here, the shapes of the initial
spectra (i.e. at hadronic freeze-out) are assumed to have been the same
for both the positive and negative pions, however the initial yields of
the two charges are different. The ratio of the initial yields is the
`initial' pion ratio $R_i$.  After freeze-out, the final-state Coulomb
interaction modifies the initial spectra, resulting in the final spectra.
The final energy ($E_f = m_t \cosh y$) is related to the initial energy
($E_i$) through the addition/subtraction of the Coulomb potential ($V_C$)
for the positive/negative pions: 
\begin{equation}
E_f = E_i \pm V_C
\end{equation}
The result is now an energy dependent final pion ratio ($R_f(E_f)$).
For a static spherical source, including the proper
Jacobian~\cite{Baym96}, $(E_f \pm V_C) \sqrt{(E_f \pm 
V_C)^2 - m^2}$, the final pion ratio as a function of $E_f$ is
given by:
\begin{equation}
R_f(E_f) = \frac{E_f-V_C}{E_f+V_C} 
\frac {\sqrt{(E_f-V_C)^2-m^2}} {\sqrt{(E_f+V_C)^2-m^2}}
\frac {n^+(E_f-V_C)} {n^-(E_f+V_C)}
\end{equation}
where the $n^{\pm}(E)$ are the pion emission functions describing the initial
$\pi^{\pm}$ spectrum. In general, the pion emission functions in heavy-ion collisions
are best represented by a Bose-Einstein distribution, so that:
\begin{equation}
\frac {n^+(E_f-V_C)} {n^-(E_i+V_C)} = \frac {A^+(e^{(E_f+V_C)/T_\pi}-1)} 
{A^-(e^{(E_f-V_C)/T_\pi} -1)}
\end{equation}
where $T_\pi$ is the slope parameter and the $A^{\pm}$ are the amplitudes
characteristic of the initial pion distributions. The initial pion ratio, 
$R_i$, is defined as $A^+/A^-$. In reference \cite{Barr00}, E877 replaced 
Eqn.~(3) with a constant `overall' pion ratio, $R'$. If one assumes a Maxwell-Boltzmann
distribution for the initial pion spectra, then $R'=R_i e^{2V_C/T_\pi}$.
However, using a Bose-Einstein form for the initial pion spectra results in
an energy dependent emission function ratio, $n^+(E_f-V_C)/n^-(E_i+V_C)$,
that can not be approximated with a constant $R'$.

The assumption of a static source is not valid for heavy-ion collisions.
During the course of the interaction, the protons, which carry the bulk of
the source charge, are emitted simultaneously with the pions. Thus the
charged source is expanding during the course of the Coulomb interaction.
Therefore, the low momentum pions do not experience the full Coulomb 
potential but rather a reduced potential. This reduced Coulomb potential, 
as a function of pion momentum, can be calculated by integrating the
proton emission function up to a maximum kinetic energy corresponding to
the pion velocity, $ E_{\rm{max}} = \sqrt{(m_p p_\pi / m_\pi)^2 + m_p^2} -
m_p$. Assuming that the proton emission function is given by a
Maxwell-Boltzmann distribution with a characteristic slope parameter 
$T_p$, the effective Coulomb potential is:
\begin{equation}
V_{\rm{eff}} = V_C (1 - e^{-E_{\rm{max}}/T_p})
\end{equation}

The mid-rapidity pion ratios for beam energies 1 to 158 AGeV are shown in 
Fig.~2 (references to the experimental data are given in the figure caption).  
The data are fit to the ratio
function as given in Eqn.~(2). The two curves in each panel correspond to
either a fixed $V_C$ or a $V_{\rm{eff}}$ given by Eqn.~(4). For these fits,
we have fixed the slope parameters of the pion and proton initial
distributions to the values given in Table~I. The pion initial slope
parameters were fixed by simultaneously fitting the $\pi^+$ and $\pi^-$
spectra in the range $0 < m_t-m_0 < 0.5$ GeV/c$^2$ to Coulomb-modified
Bose-Einstein distributions. 
\begin{equation}
\frac{1}{2 \pi m_t} \frac{d^2N}{dydm_t}(E_f) = (E_f\mp V_{eff})\frac{\sqrt{(E_f\mp V_{eff})^2 - m^2}
A^\pm} {(e^{(E_f \mp V_{eff})/T_\pi} - 1)}
\end{equation}
where $E_f = m_t{\rm cosh}y$, $A^\pm$ is a normalization constant, and $V_{eff}$ 
is given by Eqn. (4).
These fits are shown by the solid ($\pi^-$)
and dashed ($\pi^+$) curves in Fig.~1. 
The pion slope parameters used in this analysis are lower than
those reported previously~\cite{Wagn98,Ahle96,Klay03} at each beam energy
because this analysis focuses on the $m_t-m_0$ region below 0.5 GeV/c$^2$,
whereas the published slope parameters come from fits to higher $m_t-m_0$
regions of the spectra. The proton slope parameters given in Table~I were
determined using Maxwell-Boltzmann fits to spectra data from previous 
publications~\cite{Herr96,Reis10,Klay02,Back01,E917_02,Ahle99,E802_99_2,
NA49_06,NA49_11,NA44_02}. For this analysis, the
fit range was limited to $0.25 < m_t-m_0 <1.0$ GeV/c$^2$. The slope
parameters used in this analysis are similar to those cited by the authors
of the original studies.

\begin{table}[h]
\caption{$T_{\pi}$ and $T_p$ are slope parameters describing the pion and proton
spectra. These are fixed parameters in the fits to the pion
ratio data using Eqn.~(2) with
the effective Coulomb potential given in Eqn.~(4). These fits are shown by the 
solid curves in Fig.~2. The extracted Coulomb potential ($V_C$) and initial
pion ratio ($R_i$) are tabulated for each bombarding energy.}
\label{Table1}
%\begin{tabular}{ccccc}
\begin{tabular*}{0.5\textwidth}{@{}l*{15}{@{\extracolsep{0pt
          plus12pt}}l}}
$E_{\rm{beam}}$ & $T_\pi$ & $T_p$ & $V_C|_{y=0}$  & $R_i|_{y=0}$ \\
(AGeV) & (MeV) & (MeV) & (MeV) & \\
\hline
1    &  75 & 172 & 27.8 $\pm$ 1.3 & 0.469 $\pm$ .011 \\
2    &  81 & 183 & 24.8 $\pm$ 0.9 & 0.515 $\pm$ .005 \\
4    &  86 & 203 & 21.9 $\pm$ 0.5 & 0.639 $\pm$ .004 \\
6    &  92 & 216 & 19.3 $\pm$ 0.4 & 0.694 $\pm$ .004 \\
8    &  94 & 225 & 17.5 $\pm$ 0.5 & 0.710 $\pm$ .004 \\
10.8 & 100 & 229 & 16.5 $\pm$ 4.1 & 0.749 $\pm$ .035 \\
20   & 104 & 234 & 13.3 $\pm$ 3.0 & 0.834 $\pm$ .008 \\
30   & 119 & 243 &  8.8 $\pm$ 1.5 & 0.871 $\pm$ .009 \\
42   & 125 & 252 &  8.2 $\pm$ 5.0 & 0.950 $\pm$ .050 \\
158  & 130 & 257 &  7.9 $\pm$ 0.5 & 0.930 $\pm$ .003 \\
\end{tabular*}
\end{table}

\begin{figure}[th]
\includegraphics[scale=0.4]{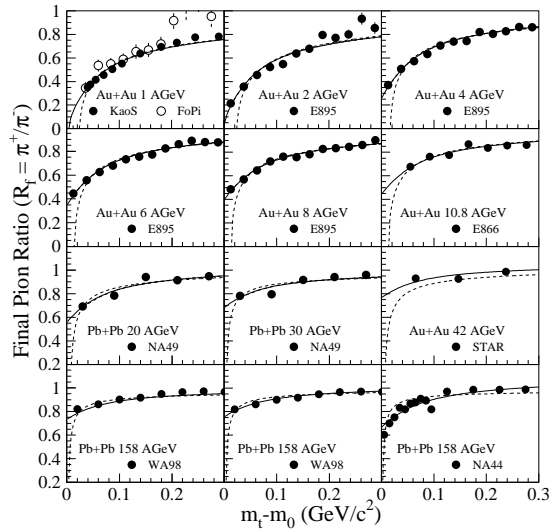}
\caption{
The mid-rapidity $\pi^+/\pi^-$ ratio as a function of
$m_t-m_0$ is shown for central Au+Au and Pb+Pb collisions at
1.0~\cite{Wagn98,Pelt97}, 2.0~\cite{Klay03}, 4.0~\cite{Klay03},
6.0~\cite{Klay03}, 8.0~\cite{Klay03}, 10.8~\cite{Ahle96}, 20~\cite{NA49_08}, 
30~\cite{NA49_08}, 42~\cite{STAR10}, and 158~\cite{Bogg96,Reti01,Ross02,NA44_02} 
AGeV. Note that there are two panels for the WA98 data as they report
results independently for the two arms of their spectrometer.
The ratios are fit with the function given in Eqn.~(2). The two
curves correspond to use of either a fixed $V_C$ (dashed) or $V_{\rm{eff}}$
as given in Eqn.~(4) (solid).}
\label{fig:y0}
\end{figure}

The fits to the pion ratio data in Fig.~2 were achieved with two free
parameters, the Coulomb potential, $V_C$, and the initial pion ratio,
$R_i$. It is evident that the low $m_t-m_0$ data points are better fit by
the effective potential (solid curve) than by a fixed $V_C$ (dashed
curve). The magnitude of the correction due to the effective Coulomb
potential of Eqn.~(4) is determined by the proton slope parameter. To test
our assumptions, we have allowed the proton slope parameter to be a third
free parameter. In these cases we found $T_p$ to be consistently  
30-50 MeV greater than the published values. This systematic discrepancy
comes from the radial flow of the protons~\cite{Schn93}. The Maxwell-Boltzmann
distribution which was used to fit the proton spectra does not
include the effect of radial flow and consistently overestimates the
proton spectra at low $m_t-m_0$. The pions are sensitive to the lower
energy range and as a result a higher proton slope parameter is suggested.  
We note that using the $V_{\rm{eff}}$ in the fits increases the $V_C$
required to match the observed results. Therefore, our extracted values of  
$V_C$ for the KaoS and E866 data are higher than those reported by their
respective collaborations~\cite{Wagn98,Ahle96,Munt98}. There is some
covariance between $V_C$ and $R_i$, thus our $R_i$ values are also
slightly lower than those reported by KaoS and E866. The $V_C$ and $R_i$
values we find using the effective Coulomb potential are reported in
Table~I. We observe a monotonic decrease in $V_C$ and a monotonic increase
in $R_i$ as $\sqrt{s_{NN}}$ increases.

\begin{figure}[th]
\includegraphics[scale=0.4]{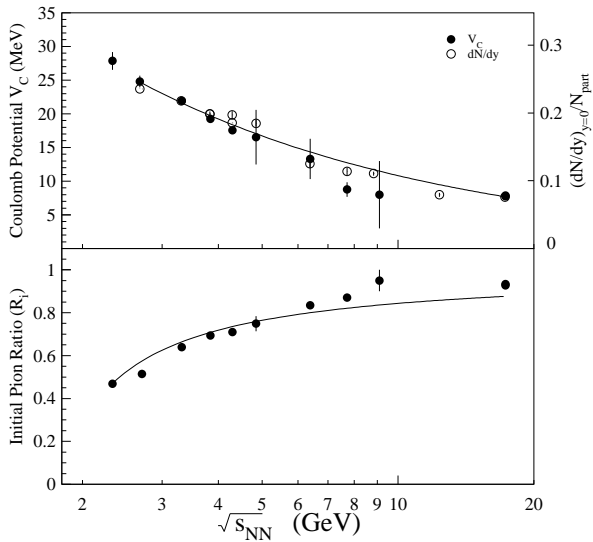}
\caption{The top panel shows in solid symbols the Coulomb potential ($V_C$) extracted from the 
fits to the pion ratio data shown in Fig.~2 as a function of center-of-mass energy 
($\sqrt{s_{NN}}$). In open symbols the normalized net-proton rapidity densities are shown, 
(d$N$/d$y$)$_{y=0}$/$N_{part}$; their axis label is on the right-hand side. The curve
in this panel is an empirical fit to the net-proton d$N$/d$y$ data. 
The proton rapidity density results are from references \cite{Klay02,E802_99_2,E917_02,NA49_06,STAR10}.
The lower panel shows the initial pion ratio, ($R_i$), extracted from the pion ratio fits as a 
function of center-of-mass energy. The fit to the data illustrates a smooth transition
from pion production exclusively through the $\Delta$ resonance channel at the lowest
collision energies to thermal production at the highest energies.}
\label{fig:3}
\end{figure}

The total charge of the interaction region is determined by the number of
participating protons and the degree of stopping for a given collision energy. 
All of the data considered are either from central Au+Au collisions or central
Pb+Pb collisions, therefore the data sets correspond to pion emission from 
sources with similar number of participating nucleons (estimates range from 
312~\cite{STAR10} to 366~\cite{NA49_06} participating nucleons) and similar 
initial overlap volume. With a static source, the Coulomb potential is determined 
by the source charge distribution and the emission point of the pion. In this
simplistic model, the monotonic decrease in $V_C$ seen in the top panel of 
Fig.~3 would correspond to an increase in the emission radius with increasing 
beam energy or a reduction in the net charge of the equilibrated system. For all beam
energies, the interaction region is first defined by the overlap of the
two colliding nuclei. A larger source size would imply that there had to
have been a period of expansion of the source prior to freeze-out, which
negates the overly simplistic static model. Indeed, there is much
evidence that heavy-ion collisions create an expanding source which can be
characterized by both radial and longitudinal flow velocities. The 
reduction in $V_C$ with beam energy is related to changes in both the size
and the shape of the charge distribution at freeze-out. However, the reduction
in the $V_C$ could also indicate that there was less net-charge in the 
interaction region due to a reduction in the baryon stopping with increased beam 
energy. To address this point, we have displayed the mid-rapidity net-proton $dN/dy$ 
values scaled by the number of participating nucleons in the same top panel of
Fig.~3 which also shows the $V_C$ values. Since we are concerned with the charge of
the interaction region, we subtract the anti-proton d$N$/d$y$ from 
proton d$N$/d$y$ to get the net-proton values. These net-proton d$N$/d$y$ values are
empirically fit with an exponential function which roughly describes the trend.
The $V_C$ values track the decrease in net-proton $dN/dy$. This suggests that the
$V_C$ is primarily measuring the charge of the net-charge of the interaction region
and the implication is that the emission radius remains unchanged across the range
of bombarding energies considered. This final conclusion is consistent with the 
trends observed for the sidewards pion source radius ($R_{side}$) as measured in
two-pion femtoscopy. In those femtoscopy studies, $R_{side}$ is observed to be 
approximately 5 fm for all $\sqrt{s_{NN}}$ from 2.5 to 200 GeV~\cite{Anson2014}.

The monotonic increase in the initial pion ratio, $R_i$, with beam energy
seen in the lower panel of Fig.~3 suggests a change in the pion production 
mechanism. At the lowest reported energies, $R_i$ is slightly below 0.5. 
This ratio value is expected if all pions were created in first-chance nucleon-nucleon 
collisions that proceeded through the $\Delta$ resonance. This value is determined 
by the neutron-to-proton ratio in the the central interaction regions of Au+Au 
collisions. From the numbers of participating protons and neutrons, the relative 
numbers of $pp$, $np$, and $nn$ collisions can be calculated. As the cross 
sections for the various $N+N \rightarrow N + N'+\pi$ channels are known~\cite{VerWest82},
one can calculate the expected $\pi^+/\pi^-$ ratio. From Glauber Monte Carlo 
models using $\sigma_{NN}$ = 45 mb (which is applicable for center-of-mass 
energies around 2.5 GeV), we estimate that for the 0-5\% central Au+Au data there 
should be 136 and 218 participating protons and neutrons respectively. Using the pion
production cross sections from \cite{VerWest82}, we estimate a $\pi^+/\pi^-$ ratio of 
0.47. It is also possible to estimate the relative yields of $\pi^+$ and $\pi^-$, 
assuming that all pions are 
created though an intermediary $\Delta$ resonance. Using the isospin of the initial 
and final states, one can calculate the relative production ratios for the various
charge states of the $\Delta$ resonance and the relative decay ratios of the $\Delta^{+}$ 
and the $\Delta^0$. Using the same number of participating protons and neutrons indicated
above and the analysis which exclusively requires production through $\Delta$ resonance
channel in first chance collisions, we expect an $R_i$ of 0.46. Both of these methodologies
reproduce the $R_i$ value extracted using the pion spectra from SIS~\cite{Wagn98,Reis07} at 
1 AGeV ($\sqrt{s_{NN}}$ = 2.33) suggesting that $\pi$ production proceeds primarily
through the $\Delta$ resonance at this energy.

The increase in $R_i$ with beam energy suggests that an increasingly larger fraction of
pions are formed in isospin independent direct production. Direct production of pion pairs 
would lead to an equal number of $\pi^+$ and $\pi^-$. This conjecture is illustrated by 
the curve in the lower panel of Fig.~3. The curve assumes that the cross sections for 
$\pi$ production through the $\Delta$ channel remain unchanged with $\sqrt{s_{NN}}$. 
For center-of-mass energies above 2.33 GeV, the cross section for $\pi$ pair production
($\sigma_{NN \rightarrow NN\pi^+\pi^-}$)
is linearly proportional to ($\sqrt{s_{NN}}$-2.33) GeV. The 
functional form of the curve is given by:
\begin{equation}
f(x) = \frac{0.47 + A (log(x)-log(2.33))}{1.0 + A(log(x)-log(2.33))}
\end{equation}
where the numerator represents the yield $\pi^+$, the denominator the yield of $\pi^-$, 
$x$ is the $\sqrt{s_{NN}}$, and $A$ is the slope of the isospin dependent pion production.
The qualitative agreement between the curve and the observed $R_i$ values suggests that
production through isospin independent channels becomes increasingly important with
collision energy.

\section{Conclusions}
We have presented a comprehensive study of the low $m_t-m_0$ pion ratios
for central Au+Au and Pb+Pb collisions from 1 to 158 AGeV. The spectra and 
ratios are fit using a model that accounts for the Coulomb potential of the source
including the effects of the pion emission functions and the radial expansion of
the source. The addition of the expansion term gives an effective Coulomb
potential which improves the fit for the lowest $m_t-m_0$ data points.
At mid-rapidity, $V_C$ falls and $R_i$ rises monotonically with bombarding
energy. The Coulomb potential is found to be consistent with a reduction in
the net-charge of the source due to the reduction in stopping as the beam energy
is increased, while the rise in the initial pion ratio suggests that the isospin 
effects become less important.

This work was supported in part
by the US National Science Foundation under Grant
No. PHY-1068833.

\bibliography{biblio_coulomb2014}

\end{document}